\documentclass[pdftex,twocolumn,epjc3,preprintnumbers,amsmath,amssymb]{svjour3}          

\RequirePackage[T1]{fontenc}

\smartqed  

\RequirePackage{graphicx}
\RequirePackage{mathptmx}   
\RequirePackage{flushend}
\RequirePackage[numbers,sort&compress]{natbib}
\RequirePackage[colorlinks,citecolor=blue,urlcolor=blue,linkcolor=blue]{hyperref}

\journalname{Eur. Phys. J. C}
\usepackage{amsmath}
\usepackage{amssymb}
\usepackage[normalem]{ulem}
\usepackage{bm}
\usepackage{color}
\newcommand{\n}{\nonumber}

\begin{document}

\title{An Exploration of the Black Hole Entropy via the Weyl Tensor} 


\author{Nan Li\thanksref{e3,addr1}
        \and
        Xiao-Long Li\thanksref{addr2}
        \and
        Shu-Peng Song\thanksref{addr3}}

\thankstext{e3}{E-mail: linan@mail.neu.edu.cn}

\institute{Department of Physics, College of Sciences, Northeastern University, Shenyang 110819, China\label{addr1}
          \and
          Department of Astronomy, Beijing Normal University, Beijing 100875, China\label{addr2}
          \and
          Department of Physics, Beijing Normal University, Beijing 100875, China\label{addr3}
}

\date{Received: date / Accepted: date}

\maketitle

\begin{abstract}
The Weyl tensor is the trace-free part of the Riemann tensor. Therefore, it is independent of the energy-momentum tensor and is thus not linked to the dynamics of gravitational fields. In this paper, we explore its possible thermodynamical property (i.e. its relationship with the black hole entropy). For a Schwarzschild black hole, the Weyl scalar invariant, $C_{\mu\nu\lambda\rho}C^{\mu\nu\lambda\rho}$, is proportional to its Bekenstein--Hawking entropy. This observation inspires us to interpret $C_{\mu\nu\lambda\rho}C^{\mu\nu\lambda\rho}$ as the entropy density of the gravitational fields of black holes. A dimensional analysis indicates that this interpretation is only valid in 5-dimensional space-time. We calculate the volume integrals of $C_{\mu\nu\lambda\rho}C^{\mu\nu\lambda\rho}$ for the 5-dimensional Schwarzschild and Schwarzschild--anti-de Sitter black holes, and find that these integrals indeed lead to the correct entropy formulae, only up to some coefficients.

\end{abstract}

\section{Introduction}

From a mathematical point of view, the curvature of a manifold is measured by the 4th-order Riemann tensor $R_{\mu\nu\lambda\rho}$. In an $n$-dimensional space-time, the Riemann tensor can be decomposed into the Ricci sector and the Weyl sector,
\begin{align}
R_{\mu\nu\lambda\rho}=&\frac{1}{n-2}(g_{\mu\lambda}R_{\nu\rho}+g_{\nu\rho}R_{\mu\lambda}-g_{\nu\lambda}R_{\mu\rho}-g_{\mu\rho}R_{\nu\lambda}) \n \\
&+\frac{1}{(n-1)(n-2)}(g_{\mu\rho}g_{\nu\lambda}-g_{\mu\lambda}g_{\nu\rho})R+C_{\mu\nu\lambda\rho}, \label{W}
\end{align}
with $R_{\mu\nu}$ being the 2nd-order Ricci tensor, $R$ the Ricci scalar, and $C_{\mu\nu\lambda\rho}$ the 4th-order Weyl tensor (namely, {\rm Riemann} $ = $ {\rm Ricci} $+$ {\rm Weyl}). However, in the theory of general relativity, the Einstein equations only associate the 2nd-order Ricci tensor $R_{\mu\nu}$ with the energy-momentum tensor. Therefore, a natural question is why the information encoded in the Weyl tensor is absent in general relativity. In other words, what is the role of the Weyl tensor in the theory of gravitation?

The Weyl tensor may be considered as a part of the Riemann tensor, containing the components not captured by the Ricci tensor. Thus, the Weyl tensor is locally independent of the energy-momentum tensor, so it may be viewed as a purely geometrical description of the curvature of a space-time. As a consequence, the Weyl tensor is linked not to the dynamical, but possibly to the thermodynamical aspects of gravitational fields. One of the exploration in this direction is the ``Weyl curvature conjecture'', or the so-called ``Penrose conjecture'' \cite{Penrose}.

Before explaining the Penrose conjecture, we briefly list the mathematical properties of the Weyl tensor. From Eq. (\ref{W}), the Weyl tensor shares the same symmetries as the Riemann tensor: $C_{\mu\nu\lambda\rho}=-C_{\nu\mu\lambda\rho}=-C_{\mu\nu\rho\lambda}$, $C_{\mu\nu\lambda\rho}=C_{\lambda\rho\mu\nu}$, and $C_{\mu\nu\lambda\rho}+C_{\mu\lambda\rho\nu}+C_{\mu\rho\nu\lambda}=0$. Also, the Weyl tensor is traceless: ${C^\lambda}_{\mu\lambda\nu}=0$. Therefore, the full contraction of the Weyl tensor, $C_{\mu\nu\lambda\rho}C^{\mu\nu\lambda\rho}$, is the principal scalar invariant that one can construct. In addition, an important theorem on the Weyl tensor is that the metric of a space-time is comformally flat, if its Weyl tensor vanishes \cite{Dabrowski:2008kx}.

From these properties, Penrose conjectured that some scalar invariant of the Weyl tensor (e.g. $C_{\mu\nu\lambda\rho}C^{\mu\nu\lambda\rho}$) is a monotonically growing function of time and could thus be identified with the gravitational entropy of the universe. This conjecture can be understood in the following way. The Friedmann--Robertson--Walker space-time is conformally flat and thus has vanishing Weyl tensor but non-vanishing Ricci tensor, whereas a Schwarzschild black hole solution has vanishing Ricci tensor but non-vanishing Weyl tensor. During the evolution, our universe evolves from an almost homogeneous and isotropic space-time to an ensemble of randomly distributed black holes in the far future. Therefore, the Weyl tensor grows relative to the Ricci tensor in the universe. A monotonically growing function reminds us of the second law of thermodynamics, so Penrose conjectured that the Weyl tensor is somehow related to the gravitational entropy in the universe. Some early works to comprehend the Penrose conjecture in the theory of gravitation can be found in Ref. \cite{lit} and in cosmology in Ref. \cite{litc}.

However, despite various attempts, to our present knowledge, Penrose's idea has never been realized beyond qualitative arguments, or been mathematically formulated in a rigorous way. In fact, even the precise notion of entropy and its relationship to the Weyl tensor are still unclear and have to be specified. The aim of this paper is just to investigate some possible connection between them. In the following two sections, we explore the possibility to calculate the entropy of Schwarzschild and Schwarzschild--anti-de Sitter black holes via the Weyl scalar invariant respectively.

\section{Entropy of the Schwarzschild black hole from the Weyl scalar invariant}

Below, we choose the natural unit system, with $c=\hbar=k_{\rm B}=1$, but we keep the gravitational constant $G$, as we will work in different dimensional space-times, in which the gravitational constants are not the same.

The metric of a Schwarzschild black hole with mass $M$ reads
\begin{align}
ds^2=&-\left(1-\frac{R_{\rm S}}{r}\right)dt^2+\left(1-\frac{R_{\rm S}}{r}\right)^{-1}dr^2 \n\\
&+r^2(d\theta^2+\sin^2\theta\, d\phi^2), \n
\end{align}
where $R_{\rm S}=2GM $
is the Schwarzschild radius. From this metric, the corresponding Weyl scalar invariant is
\begin{align}
C_{\mu\nu\lambda\rho}C^{\mu\nu\lambda\rho}=\frac{48(GM)^2}{r^6}=\frac{12R_{\rm S}^2}{r^6}. \label{cc}
\end{align}
Meanwhile, the Bekenstein--Hawking entropy $S$ of the Sch\-warz\-schild black hole is \cite{Bekenstein}
\begin{align}
S=\frac{A}{4G}=\frac{4\pi R_{\rm S}^2}{4G}, \label{S}
\end{align}
where $A=4\pi R_{\rm S}^2$ is the area of its horizon. From Eqs. (\ref{cc}) and (\ref{S}), we clearly see that $C_{\mu\nu\lambda\rho}C^{\mu\nu\lambda\rho}$ is proportional to $S$. From this proportion, we naturally wonder if there is some latent relationship between the Weyl tensor and the entropy of black holes and gravitational fields.

We see from Eq. (\ref{cc}) that $C_{\mu\nu\lambda\rho}C^{\mu\nu\lambda\rho}$ is coordinate-dependent (i.e. it is the function of the radial coordinate $r$). This observation inspires us to interpret it as the entropy density of the gravitational field of the Schwarzschild black hole, and its volume integral may thus give the Bekenstein--Hawking entropy.

However, a further consideration easily invalidates this simple attempt. The dimension of the Weyl tensor, $[C_{\mu\nu\lambda\rho}]$, is $+2$ in the natural unit system, so the dimension of the Weyl scalar invariant, $[C_{\mu\nu\lambda\rho}C^{\mu\nu\lambda\rho}]$, should be $+4$. On the other hand, in a 4-dimensional space-time, the dimension of volume element is $-3$. Therefore, we are not allowed to expect $C_{\mu\nu\lambda\rho}C^{\mu\nu\lambda\rho}$ as the entropy density of gravitational field, as its volume integral should have dimension $+1$, but entropy itself is dimensionless in the natural unit system.

This observation enlightens us to regard $C_{\mu\nu\lambda\rho}C^{\mu\nu\lambda\rho}$ as the entropy density in a 5-dimensional space-time, in which the dimension of volume element is $-4$. Hence, the volume integral
\begin{align}
\int C_{\mu\nu\lambda\rho}C^{\mu\nu\lambda\rho} \,dV_4\label{jifen}
\end{align}
is dimensionless. We will show in the following that this integral does lead to the correct entropy formulae for the 5-dimensional Schwarzschild and Schwarzschild--anti-de Sitter black holes, only up to some coefficients.

The integral in Eq. (\ref{jifen}) consists of three parts: the integrand $C_{\mu\nu\lambda\rho}C^{\mu\nu\lambda\rho}$, the 4-dimensional invariant volume element $dV_4$, and the domain of integration. Below, we discuss them in order.

(1) For $C_{\mu\nu\lambda\rho}C^{\mu\nu\lambda\rho}$, from Eq. (\ref{W}), the explicit form of the Weyl tensor in 5-dimensional space-time is
\begin{align}
C_{\mu\nu\lambda\rho}=&R_{\mu\nu\lambda\rho}+\frac{1}{3}(g_{\nu\lambda}R_{\mu\rho}+g_{\mu\rho}R_{\nu\lambda}-g_{\mu\lambda}R_{\nu\rho} \n\\
&-g_{\nu\rho}R_{\mu\lambda})+\frac{1}{12}(g_{\nu\rho}g_{\mu\lambda}-g_{\nu\lambda}g_{\mu\rho})R. \label{5}
\end{align}
The metric of the 5-dimensional Schwarzschild black hole with mass $M$ reads \cite{Horowitz}
\begin{align}
ds^2=&-\left(1-\frac{R_5^2}{r^2}\right)dt^2+\left(1-\frac{R_5^2}{r^2}\right)^{-1}dr^2 \n\\
&+r^2(d\theta^2+\sin^2\theta\, d\phi^2+\sin^2\theta\sin^2\phi \,d\psi^2), \label{5sch}
\end{align}
where
\begin{align}
R_5=\sqrt{\frac{8G_5M}{3\pi}} \n
\end{align}
is the 5-dimensional Schwarzschild radius, and $G_5$ is the gravitational constant in 5-dimensional space-time.

From Eqs. (\ref{5}) and (\ref{5sch}), a direct calculation arrives at
\begin{align}
C_{\mu\nu\lambda\rho}C^{\mu\nu\lambda\rho}=\frac{72 R_5^4}{r^8}. \label{cccc}
\end{align}
This contraction can also be obtained via the relation,
\begin{align}
C_{\mu\nu\lambda\rho}C^{\mu\nu\lambda\rho}=&R_{\mu\nu\lambda\rho}R^{\mu\nu\lambda\rho}-\frac{4}{n-2}R_{\mu\nu}R^{\mu\nu} \n\\
&+\frac{2}{(n-1)(n-2)}R^2, \n
\end{align}
where $R_{\mu\nu\lambda\rho}R^{\mu\nu\lambda\rho}$ is the Kretschmann scalar invariant. The calculation of $C_{\mu\nu\lambda\rho}C^{\mu\nu\lambda\rho}$ in this way is much easier, since both the Ricci tensor and Ricci scalar vanish for the Schwarzschild metric.

(2) For $dV_4$, from Eq. (\ref{5sch}), we have
\begin{align}
dV_4=r^3\sqrt{|g_{rr}|}\,drd\Omega_3, \n
\end{align}
where $d\Omega_3$ is the solid angle element of the 3-dimensional sphere in 4-dimensional space. We may first integrate the angular parts, and
\begin{align}
\int C_{\mu\nu\lambda\rho}C^{\mu\nu\lambda\rho}\, dV_4&=\int C_{\mu\nu\lambda\rho}C^{\mu\nu\lambda\rho}r^3\sqrt{|g_{rr}|}\,drd\Omega_3 \n\\
&=2\pi^2\int C_{\mu\nu\lambda\rho}C^{\mu\nu\lambda\rho}r^3\sqrt{|g_{rr}|}\,dr, \label{li}
\end{align}
where $\int d\Omega_3=2\pi^2$ is the solid angle of the 3-dimensional sphere, and $\sqrt{|g_{rr}|}\,dr=|1-{R_5^2}/{r^2}|^{-1/2}\,dr$ is the proper distance element in the radial direction.

(3) For the domain of integration of the radial coordinate $r$, the upper limit can be safely set to be infinity, but the lower limit cannot be taken as 0, which diverges the integral in Eq. (\ref{jifen}). Actually, the classical theory of general relativity is invalid at extremely small radius (about the Planck length). Therefore, we first simply set the lower limit of $r$ to be the 5-dimensional Planck length $l_5$, which can be expressed in terms of the 5-dimensional gravitational constant $G_5$ as \cite{Zw}
\begin{align}
l_5=\sqrt[3]{G_5}. \n
\end{align}

With all these preparations, substituting Eq. (\ref{cccc}) into (\ref{li}), we attain the volume integral of the Weyl scalar invariant for the 5-dimensional Schwarzschild black hole,
\begin{align}
&\int C_{\mu\nu\lambda\rho}C^{\mu\nu\lambda\rho}\,dV_4
=\int^{\infty}_{l_5} \frac{72 R_5^4}{r^8}\frac{2\pi^2r^3\,dr}{\sqrt{\left|1-\frac{R_5^2}{r^2}\right|}} \n \\
=&144\pi^2 R_5^4 \left(\int^{R_5}_{l_5}\frac{dr}{r^5\sqrt{\frac{R_5^2}{r^2}-1}}+
\int^{\infty}_{R_5}\frac{dr}{r^5\sqrt{1-\frac{R_5^2}{r^2}}}\right) \n\\
=&48\pi^2\left\{\left[\left(\frac{R_5}{l_5}\right)^3+\frac{2R_5}{l_5}\right]\sqrt{1-\left(\frac{l_5}{R_5}\right)^2}+2\right\}. \n
\end{align}
This result seems lengthy at first glance, but in fact not. As the Schwarzschild radius of a typical celestial body is always much larger than the Planck length, so in the limit $R_5\gg l_5$, the above result can be significantly simplified, with the leading term being
\begin{align}
\int C_{\mu\nu\lambda\rho}C^{\mu\nu\lambda\rho}\,dV_4\approx 48\pi^2 \left(\frac{R_5}{l_5}\right)^3. \label{wuwei}
\end{align}

On the other hand, the Bekenstein--Hawking entropy for the 5-dimensional Schwarzschild black hole is
\begin{align}
S_5=\frac{A_5}{4G_5}=\frac{\pi^2}{2} \left(\frac{R_5}{l_5}\right)^3, \n
\end{align}
where $A_5=2\pi^2R_5^3$ is the area (i.e. the 3-dimensional volume) of its horizon. To this point, we eventually find that the volume integral of the Weyl scalar invariant indeed leads to the correct entropy formula, only up to a trivial numerical coefficient,
\begin{align}
S_5=\frac{1}{96}\int C_{\mu\nu\lambda\rho}C^{\mu\nu\lambda\rho}\,dV_4. \n
\end{align}
From this result, we are convinced that the interpretation of the Weyl scalar invariant $C_{\mu\nu\lambda\rho}C^{\mu\nu\lambda\rho}$ as the entropy density for the 5-dimensional Schwarzschild black hole is reasonable.

\section{Entropy of the Schwarzschild--anti-de Sitter black hole from the Weyl scalar invariant}

Till now, we only discuss the simplest Schwarzschild black hole solution, for which the Weyl tensor is identical to the Riemann tensor, $C_{\mu\nu\lambda\rho}=R_{\mu\nu\lambda\rho}$, and the Weyl scalar invariant coincides with the Kretschmann scalar invariant, so the characteristic of the Weyl tensor is not distinct. Therefore, we further explore a more complicated black hole with non-vanishing Ricci tensor and Ricci scalar: the Schwarzschild--anti-de Sitter solution (i.e. the Schwarzschild solution with a cosmological constant $\Lambda<0$). In 5-dimensional space-time, its metric reads
\begin{align}
ds^2=&-\left(1-\frac{R_5^2}{r^2}-\frac{\Lambda r^2}{6}\right)dt^2+\left(1-\frac{R_5^2}{r^2}-\frac{\Lambda r^2}{6}\right)^{-1}dr^2 \n\\
&+r^2(d\theta^2+\sin^2\theta\, d\phi^2+\sin^2\theta\sin^2\phi \,d\psi^2), \label{schds}
\end{align}

In this case, both the Ricci tensor and Ricci scalar are non-vanishing: $R_{\mu\nu}=\frac23\Lambda g_{\mu\nu}$ and $R=\frac{10}{3}\Lambda$. Therefore, the difference between the Weyl and Riemann tensors becomes much more notable. It is straightforward to see that the Kretschmann scalar invariant now receives a modification from the cosmological constant,
\begin{align}
R_{\mu\nu\lambda\rho}R^{\mu\nu\lambda\rho}=\frac{72R_5^4}{r^8}+\frac{10\Lambda^2}{9}, \label{1}
\end{align}
but the Weyl scalar invariant remains unchanged,
\begin{align}
C_{\mu\nu\lambda\rho}C^{\mu\nu\lambda\rho}=\frac{72R_5^4}{r^8}. \label{2}
\end{align}
These results strongly support our interpretation of the Weyl scalar invariant, not the Kretschmann scalar invariant, as the entropy density of gravitational field. Below, we perform the parallel procedure for the volume integral in Eq. (\ref{jifen}), in order to check the validity of our interpretation.

In the similar way, we have
\begin{align}
&\int C_{\mu\nu\lambda\rho}C^{\mu\nu\lambda\rho}\,dV_4
=\int^{\infty}_{l_5} \frac{72R_5^4}{r^8}\frac{2\pi^2r^3\,dr}{\sqrt{\left|1-\frac{R_5^2}{r^2}-\frac{\Lambda r^2}{6}\right|}} \n\\
=&144\pi^2R_5^4\left(\int^{R_5'}_{l_5}\frac{dr}{r^5\sqrt{\frac{R_5^2}{r^2}+\frac{\Lambda r^2}{6}-1}} \right.  \n\\
&\left. ~~~~~~~~~~~~~~ +\int_{R_5'}^\infty\frac{dr}{r^5\sqrt{1-\frac{R_5^2}{r^2}-\frac{\Lambda r^2}{6}}}\right), \label{l}
\end{align}
where
\begin{align}
R_5'=\sqrt{\frac{3}{\Lambda}\left(1-\sqrt{1-\frac{2\Lambda R_5^2}{3}}\right)} \label{r'}
\end{align}
is the unique zero for $1-{R_5^2}/{r^2}-{\Lambda r^2}/{6}$ (since $\Lambda<0$).

The results of the integrals in Eq. (\ref{l}) can be expressed analytically via the elliptic functions, which depend on $R_5$ and $\Lambda$. However, the exact but tedious expressions are irrelevant, but only the leading term is important,
\begin{align}
\int C_{\mu\nu\lambda\rho}C^{\mu\nu\lambda\rho}\,dV_4\approx 48\pi^2 \left(\frac{R_5}{l_5}\right)^3. \n
\end{align}
This result is the same as that in Eq. (\ref{wuwei}), indicating that the cosmological constant does not significantly affect the integral in Eq. (\ref{l}). This fact is not difficult to understand, as the cosmological constant contributes to the integral only at large $r$, where its effect is dominantly suppressed by the factor $1/r^8$ in the Weyl scalar invariant, so we arrive at the same result in Eq. (\ref{wuwei}).

Meanwhile, the Bekenstein--Hawking entropy for the 5-dimensional Schwarzschild--anti-de Sitter black hole is \cite{page}
\begin{align}
S_5=\frac{A_5'}{4G_5}=\frac{\pi^2}{2} \left(\frac{R_5'}{l_5}\right)^3, \n
\end{align}
where $A_5'=2\pi^2{R'_5}^3$ is the area (i.e. the 3-dimensio\-nal volume) of its horizon. Thus, we have
\begin{align}
S_5'=\frac{1}{96}\left(\frac{R_5'}{R_5}\right)^3\int C_{\mu\nu\lambda\rho}C^{\mu\nu\lambda\rho}\,dV_4. \n
\end{align}
Hence, we again obtain the entropy formula, but this time the coefficient is $\Lambda$-dependent. From Eq. (\ref{r'}), for a small negative $\Lambda$, $({R_5'}/{R_5})^3\rightarrow 1$; for a large negative $\Lambda$, $({R_5'}/{R_5})^3\rightarrow [6/(-\Lambda R_5^2)]^{3/4}$. Although the coefficient varies with $\Lambda$, the proportion remains the same.

In short, for the 5-dimensional Schwarzschild--anti-de Sitter black hole, for which the Weyl tensor deviates from the Riemann tensor, $C_{\mu\nu\lambda\rho}\neq R_{\mu\nu\lambda\rho}$, we are still able to obtain the Bekenstein--Hawking entropy by integrating the Weyl scalar invariant $C_{\mu\nu\lambda\rho}C^{\mu\nu\lambda\rho}$, as the effect of the cosmological constant is suppressed at large radius. Especially, from Eqs. (\ref{1}) and (\ref{2}), we explicitly see the difference between the Kretschmann scalar invariant $R_{\mu\nu\lambda\rho}R^{\mu\nu\lambda\rho}$ and the Weyl scalar invariant $C_{\mu\nu\lambda\rho}C^{\mu\nu\lambda\rho}$. We find that $C_{\mu\nu\lambda\rho}C^{\mu\nu\lambda\rho}$ is not altered in the presence of the cosmological constant, and this fact strengthens our interpretation of the Weyl scalar invariant as the entropy density, but not the Kretschmann scalar invariant, since it receives a constant modification $10\Lambda^2/9$ from the cosmological constant, which diverges the volume integral when $r$ goes to infinity.

Here, we should also mention that the thermodynamics of the Schwarzschild--de Sitter black hole (i.e. the Sch\-warz\-schild solution with a positive cosmological constant) is not a well-defined issue \cite{sds}, and we skip the corresponding discussion in the present paper.

\section{Conclusions and discussions}

Finally, we give some brief discussions on our work. The issue of entropy is one of the central problems in black hole thermodynamics \cite{entropy}, and the Penrose conjecture on the Weyl tensor is one of the possible ways to approach this issue. We try in this paper to physically confirm and mathematically formulate the Penrose conjecture, i.e. we explore the latent relationship between the Weyl tensor and the entropy of gravitational fields. We take the Schwarzschild and Schwarzschild--anti-de Sitter black holes as examples and find that we may interpret the Weyl scalar invariant $C_{\mu\nu\lambda\rho}C^{\mu\nu\lambda\rho}$ as their entropy densities, but this should be realized in 5-dimensional space-time from the dimensional analysis. We perform the volume integral of $C_{\mu\nu\lambda\rho}C^{\mu\nu\lambda\rho}$ from the 5-dimensional Planck length to infinity, and discover that this volume integral really results in the correct Bekenstein--Hawking entropy formulae, only up to some coefficients.

At the same time, we should also point out the limits of our work. First, our calculation only applies in 5-dimensional space-time, but not the 4-dimensional space-time that we live in. We may otherwise imagine that the mass of the 5-dimensional black hole is distributed on an extra dimension, and if this dimension is wrapped to an extremely small scale, we can still utilize our method to calculate the volume integral of the corresponding Weyl scalar invariant. But in this case, what we face is to calculate the metric and the Weyl tensor of a black string, and this calculation is much more complicated and is thus beyond our preliminary exploration. (For the metric solution for a black string, see Ref. \cite{blackstring}.) Second, we should admit that it is difficult to extend our results to the more general black holes (e.g. the charged Reissner--Nordstr\"{o}m black hole), because its geometry near the origin is quite different from that of the Schwarzschild and Schwarzschild--anti-de Sitter black holes.

In summary, our work helps to understand Penrose's idea and indicates that the Weyl tensor may be related to the entropy of gravitational fields, but some difficulties are still to be overcome. This exploration leads us to investigate whether there exist equations that are parallel to the Einstein equations and quantify the thermodynamical relationship between space-time and matter. These equations are expected to relate the Weyl tensor to the thermodynamical concepts such as entropy and temperature, and we wish that our work would be conducive to the research in this direction in future.

\vskip .3cm

We are very grateful to Tian-Fu Fu, Xiao-Ran Han, and Dominik J. Schwarz for fruitful discussions. This work is supported by the Fundamental Research Funds for the Central Universities of China (No. N140504008).

\end{document}